\documentclass[sigconf]{acmart}
\AtBeginDocument{%
  }

\usepackage{bm}
\usepackage{float}
\usepackage{tabularx}
\usepackage{subcaption}

\setcopyright{acmlicensed}
\copyrightyear{2026}
\acmYear{2026}
\acmConference[ICAIF '26]{6th ACM International Conference on AI in Finance}{November 14--17, 2026}{Milan, Italy}

\begin{document}

\title{Universal Diffusion Models for Implied Volatility Surfaces: Learning Shared Dynamics Across Stocks}

\author{Mingzhi Yang}
\email{myang682@connect.hkust-gz.edu.cn}

\author{Sheng Wang}
\email{victorshengw@hkust-gz.edu.cn}

\author{Chao Zhang}
\correspondingauthor
\email{chaoz@hkust-gz.edu.cn}

\author{Ruikun Li}
\email{ruikunli@hkust-gz.edu.cn}
\affiliation{%
  \institution{The Hong Kong University of Science and Technology (Guangzhou)}
  \city{Guangzhou}
  \country{China}
}

\renewcommand{\shortauthors}{Mingzhi et al.}

\begin{abstract}
Modeling the dynamics of option implied volatility surface (IVS) is crucial for pricing, hedging, and risk-managing option portfolios. We develop a universal conditional diffusion model that learns to jointly generate next-day IVS increments and the underlying stock's returns. The model is trained on pooled data from 50 stocks and evaluated on a test set comprising 50 in-sample stocks and 50 out-of-sample stocks excluded from training. The training objective is primarily the minimization of MSE, with the variants that jointly impose surface smoothness, and that penalize the presence of static-arbitrage, which are all economically meaningful constraints. Across in-sample and out-of-sample stocks, our diffusion model outperforms the VolGAN \cite{VuleticCont2024} benchmark in reducing arbitrage violations, improving stock risk prediction, and aligning explained variance ratios by the first three principal components. The fact that the model extrapolates well to stocks that are excluded from training indicates that the learned dynamics can be shared across stocks. These findings support a universal conditional diffusion as a credible framework for multi-stock implied volatility surface scenario generation.
\end{abstract}




\maketitle

\section{Introduction}

The implied volatility surface (IVS) of an option, recovered by inverting the Black--Scholes formula~\cite{BlackScholes1973}, provides a standardized volatility metric across strikes and maturities, exhibiting smile, skew, and term-structure patterns~\cite{ContdaFonseca2002,Gatheral2006}. Modeling its dynamics is challenging due to sparse quotes, high dimensionality, and no-arbitrage constraints~\cite{Fengler2009,GatheralJacquier2014}. Classical structural models (e.g., deterministic-volatility, implied-tree, stochastic volatility, and jump models)~\cite{Rubinstein1994,DumasFlemingWhaley1998,Heston1993,Bates1996,BakshiCaoChen1997,ChristoffersenHestonJacobs2009} provide economic interpretability but are limited by pre-specified dynamics. Data-driven generative models offer greater flexibility. Generative adversarial networks (GANs)~\cite{Goodfellow2014GAN} have been applied to financial time series~\cite{Karras2019StyleGAN,Yoon2019TimeGAN,Wiese2020QuantGAN}, and VolGAN~\cite{VuleticCont2024} uses a conditional GAN for joint return and IVS change generation with smoothness regularization.

Diffusion models~\cite{Ho2020DDPM,Song2021SDE} learn distributions via denoising, achieving state-of-the-art results across image, audio, video, text, molecular, robotics, and weather domains~\cite{DhariwalNichol2021,Kong2021DiffWave,Ho2022VideoDiffusion,Li2022DiffusionLM,Hoogeboom2022Molecule,Chi2023DiffusionPolicy,Price2025GenCast}, and have been adapted for probabilistic time-series forecasting~\cite{Rasul2021TimeGrad,Tashiro2021CSDI,ShenKwok2023TimeDiff}. Recent finance-specific diffusion models show improved fidelity over GAN-based baselines~\cite{HuangChenQiao2024FTSDiffusion,TakahashiMizuno2025}. Complementary evidence from limit-order-book data suggests that deep learning models trained on pooled cross-stock data can capture universal price-formation features and generalize to stocks excluded from training~\cite{SirignanoCont2019}. Most closely related to our setting, Han et al.~\cite{Han2026ADSeqVol} develop an adaptive sequential diffusion model that jointly generates SPX returns and implied-volatility surfaces conditional on recent market history, and incorporate static no-arbitrage constraints through post-training fine-tuning. We instead study whether a single diffusion model trained on a pooled cross-section of equities can learn IVS dynamics that transfer to previously unseen stocks without retraining.

Our contribution is a universal conditional diffusion framework for IVS dynamics. We pool observations from 50 stocks to train a single model that jointly generates the next-day return and IVS increment, with static-arbitrage control during training and generation. We evaluate the model's transfer to 50 out-of-sample stocks without retraining, testing whether the learned dynamics are shared across stocks. The model architecture uses a FiLM-modulated denoiser, and we compare variants with smoothness, arbitrage, and hybrid losses against a pooled VolGAN benchmark.

\section{Implied Volatility Surface}\label{sec:ivs_foundations}

Options are typically quoted in terms of implied volatilities. These volatilities are recovered from option premiums by inverting the Black–Scholes formula \cite{BlackScholes1973,Merton1973}. Given the date $t$ and underlying stock price $S_t$, for an option with strike price $K$ and maturity date $T$, we define its \emph{moneyness} as $m = K/S_t$ and \emph{time-to-maturity} as $\tau = T - t$. The IVS of the underlying at the date $t$ is given by the map $(m,\tau) \mapsto \sigma_t(m,\tau)$, whose departures from flatness are known as the stylized smile, skew, and term-structure patterns \cite{ContdaFonseca2002}. The surface's arbitrage-free structure is governed by shape restrictions on the resulting option-price surface that rule out static arbitrage \cite{Fengler2009,GatheralJacquier2014}.

\subsection{Arbitrage Penalties on a Discrete Grid}

On the discrete grid, static arbitrage constraints are enforced via finite-difference penalties~\cite{ContVuletic2023}. Let $\bm{m} = (m_1, \dots, m_{N_m})$ and $\bm{\tau} = (\tau_1, \dots, \tau_{N_\tau})$ be ordered moneyness and maturity nodes. The total violation is
\begin{equation}\label{eq:arbpenalty}
\begin{split}
    \Phi\big(\sigma(\bm{m}, \bm{\tau})\big) 
    &= p_{\mathrm{calendar}}\big(\sigma(\bm{m}, \bm{\tau})\big) \\
    &\ + p_{\mathrm{spread}}\big(\sigma(\bm{m}, \bm{\tau})\big) 
     + p_{\mathrm{butterfly}}\big(\sigma(\bm{m}, \bm{\tau})\big).
\end{split}
\end{equation}
where the components are given by
\begin{equation}\label{eq:p1}
    p_{\mathrm{calendar}}\big(\sigma(\bm{m}, \bm{\tau})\big) = \sum_{i=1}^{N_m} \sum_{j=1}^{N_\tau-1} \left( \tau_j \frac{c(m_i, \tau_j) - c(m_i, \tau_{j+1})}{\tau_{j+1} - \tau_j} \right)^+,
\end{equation}
\begin{equation}\label{eq:p2}
    p_{\mathrm{spread}}\big(\sigma(\bm{m}, \bm{\tau})\big) = \sum_{i=1}^{N_m-1} \sum_{j=1}^{N_\tau} \left( \frac{c(m_{i+1}, \tau_j) - c(m_i, \tau_j)}{m_{i+1} - m_i} \right)^+.
\end{equation}
\begin{equation}\label{eq:p3}
\begin{split}
    p_{\mathrm{butterfly}}\big(\sigma(\bm{m}, \bm{\tau})\big)
    &= \sum_{i=2}^{N_m-1} \sum_{j=1}^{N_\tau}
    \Bigg( \frac{c(m_i, \tau_j) - c(m_{i-1}, \tau_j)}{m_i - m_{i-1}} \\
    &\qquad - \frac{c(m_{i+1}, \tau_j) - c(m_i, \tau_j)}{m_{i+1} - m_i} \Bigg)^+.
\end{split}
\end{equation}
The arbitrage-free condition is $\Phi\big(\sigma(\bm{m}, \bm{\tau})\big) = 0$.

\subsection{Smoothing the IVS}

\begin{figure}[!ht]
    \centering

    \begin{subfigure}[t]{0.48\linewidth}
        \centering
        \includegraphics[width=\linewidth]{%
            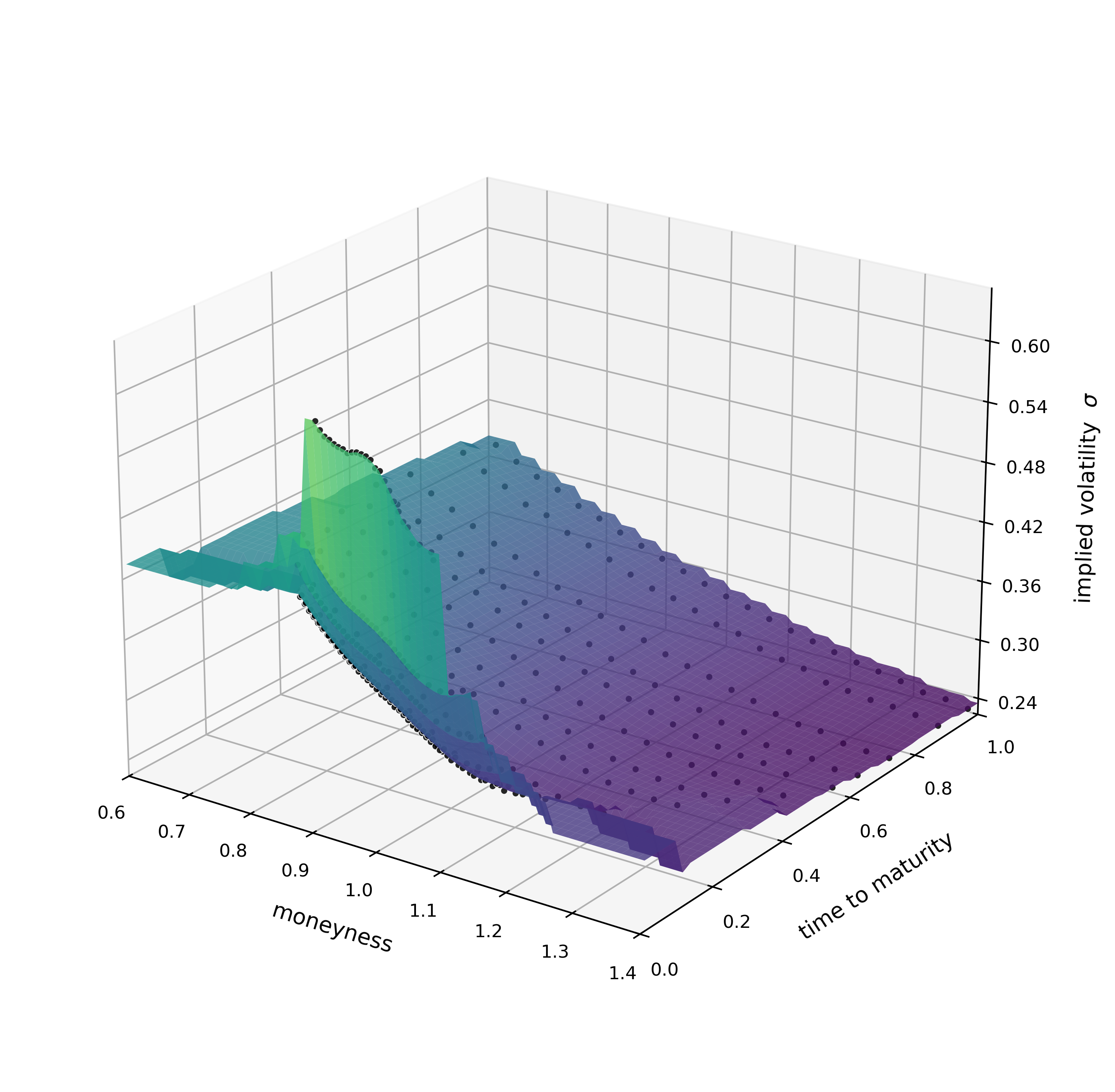}
        \caption{Raw option quotes.}
        \label{fig:raw}
    \end{subfigure}
    \hfill
    \begin{subfigure}[t]{0.48\linewidth}
        \centering
        \includegraphics[width=\linewidth]{%
            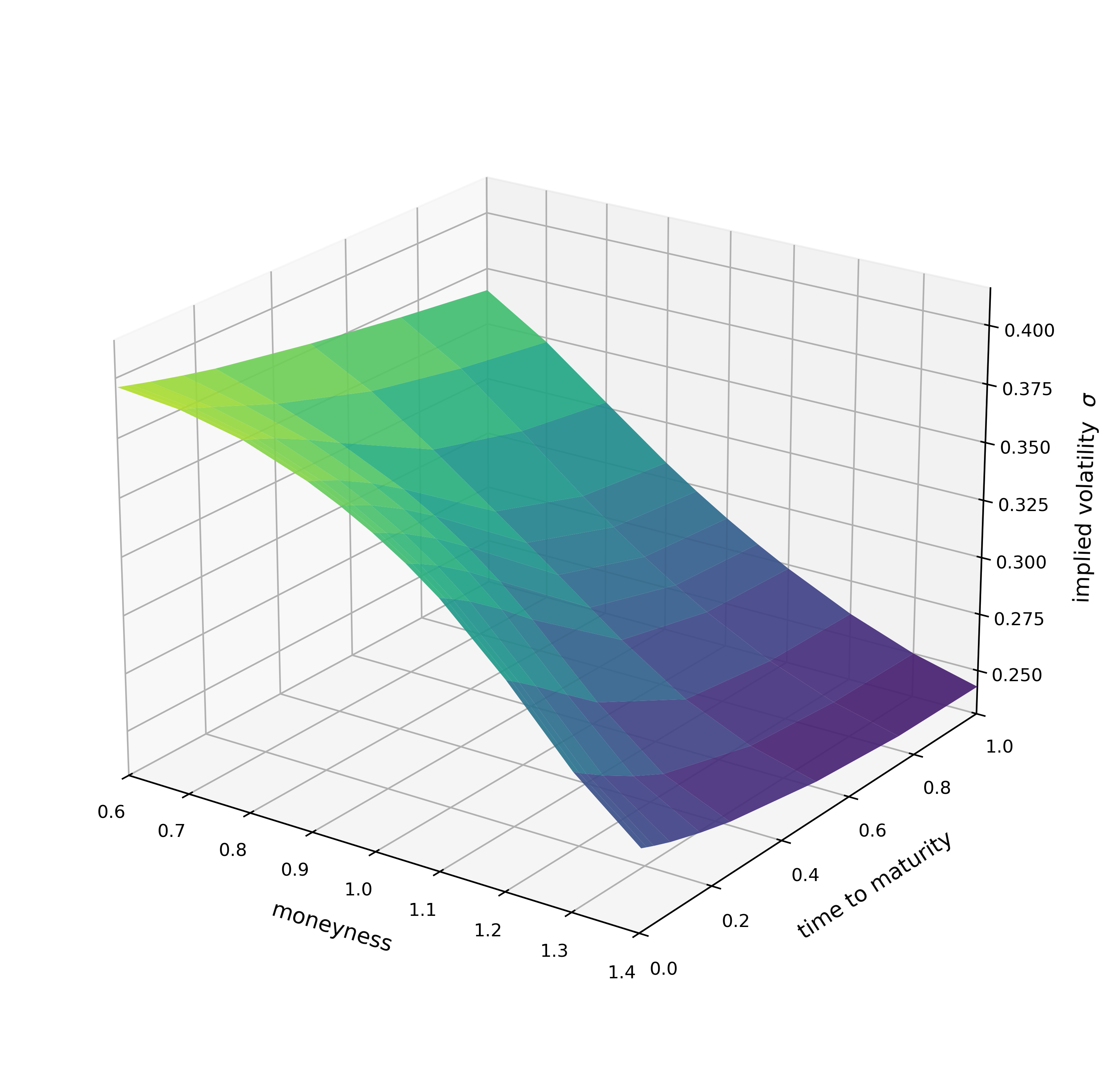}
        \caption{Smoothed IV surface.}
        \label{fig:smoothed}
    \end{subfigure}

    \caption{
        Transformation of irregular option quotes (a) into a fixed-dimensional implied-volatility surface (b). The sample is based on AAPL option quotes observed on August 27, 2024.
    }
    \label{fig:smoothing_pipeline}
\end{figure}

Raw quotes are filtered, restricted to out-of-the-money contracts, and converted to a fixed-dimensional surface representation~\cite{ContdaFonseca2002,Fengler2009,VuleticCont2024}. We adopt a common grid with moneyness nodes
\begin{equation}\label{eq:moneygrid}
    \mathcal{M} = \{0.6,\,0.7,\,0.8,\,0.9,\,0.95,\,1,\,1.05,\,1.1,\,1.2,\,1.3,\,1.4\},
\end{equation}
denser near the money, and maturity nodes (in years)
\begin{equation}\label{eq:taugrid}
    \mathcal{T} = \Big\{\tfrac{1}{252},\,\tfrac{1}{52},\,\tfrac{2}{52},\,\tfrac{1}{12},\,\tfrac{1}{6},\,\tfrac{1}{4},\,\tfrac{1}{2},\,\tfrac{3}{4},\,1\Big\}.
\end{equation}
A Vega-weighted Nadaraya--Watson smoother with stock-specific bandwidths (tuned to minimize static arbitrage) is applied to observed quotes and then interpolated onto $\mathcal{T}$, producing an $11\times 9$ log-volatility matrix flattened into a 99-dimensional vector. Figure~\ref{fig:smoothing_pipeline} illustrates the transformation.

\section{Diffusion Models for IVS}\label{sec:diffusion_models}

The generative model is a single conditional denoising diffusion network shared across the pooled cross-section of stocks. Diffusion models have recently been applied to dynamic IVS generation, including joint modeling of underlying returns and volatility surfaces~\cite{Han2026ADSeqVol}. To keep the comparison with VolGAN disciplined, our state definition and target construction follow the same economic decomposition used in VolGAN~\cite{VuleticCont2024}, with the model conditioning on the recent market state and generating the next-day return together with the next-day surface increment. Arbitrage control enters both the training objective and the generation stage~\cite{Ho2020DDPM,ContVuletic2023,VuleticCont2024,Han2026ADSeqVol}.

\subsection{Variables}

Following the inputs and outputs of VolGAN \cite{VuleticCont2024}, let $R_t$ denote the underlying log-return and let $g_t = \log\sigma_t$ be the 99-dimensional log-volatility vector obtained from the grid-aligned surface. The conditioning state at the date $t$ is
\begin{equation}\label{eq:cond}
    c_t = \big[ R_{t-1}, R_{t-2}, RV_{t-1}, g_{t-1} \big] \in \mathbb{R}^{102},
\end{equation}
where $R_{t-1}$ and $R_{t-2}$ denote the two lagged daily log-returns, $RV_{t-1}$ denotes monthly realized volatility, given by
\begin{equation}\label{eq:rv}
    RV_{t-1} = \sqrt{\frac{252}{21} \sum_{j=1}^{21} R_{t-j}^{2}}.
\end{equation}
The model learns to generate the joint next-day innovation
\begin{equation}\label{eq:target}
    x_t = \big[ R_t, \Delta g_t \big] \in \mathbb{R}^{100}, \qquad \Delta g_t = g_t - g_{t-1}.
\end{equation}
This formulation anchors every synthetic scenario to the observed previous-day surface while preserving the dependence between spot returns and surface dynamics.

\subsection{Architecture}
The denoiser in the diffusion network is a FiLM-modulated multilayer perceptron shared across stocks \cite{Perez2018FiLM}. Given a noised target $x_k$, a diffusion step $k$, and the condition $c_t$, the context modulates each residual block through
\begin{equation}\label{eq:film}
    \widetilde{h}_\ell = (1 + \gamma_\ell) \odot \mathrm{LN}(h_\ell) + \beta_\ell,
\end{equation}
where $h_\ell$ denotes the hidden state in residual block $\ell$, $\mathrm{LN}(\cdot)$ denotes layer normalization, $\gamma_\ell$ and $\beta_\ell$ denote FiLM scale and shift parameters generated from the conditioning state, and $\odot$ denotes elementwise multiplication. Thus, the same target is denoised differently across market regimes and diffusion steps.

\begin{figure}[H]
\centering
\includegraphics[width=0.49\textwidth]{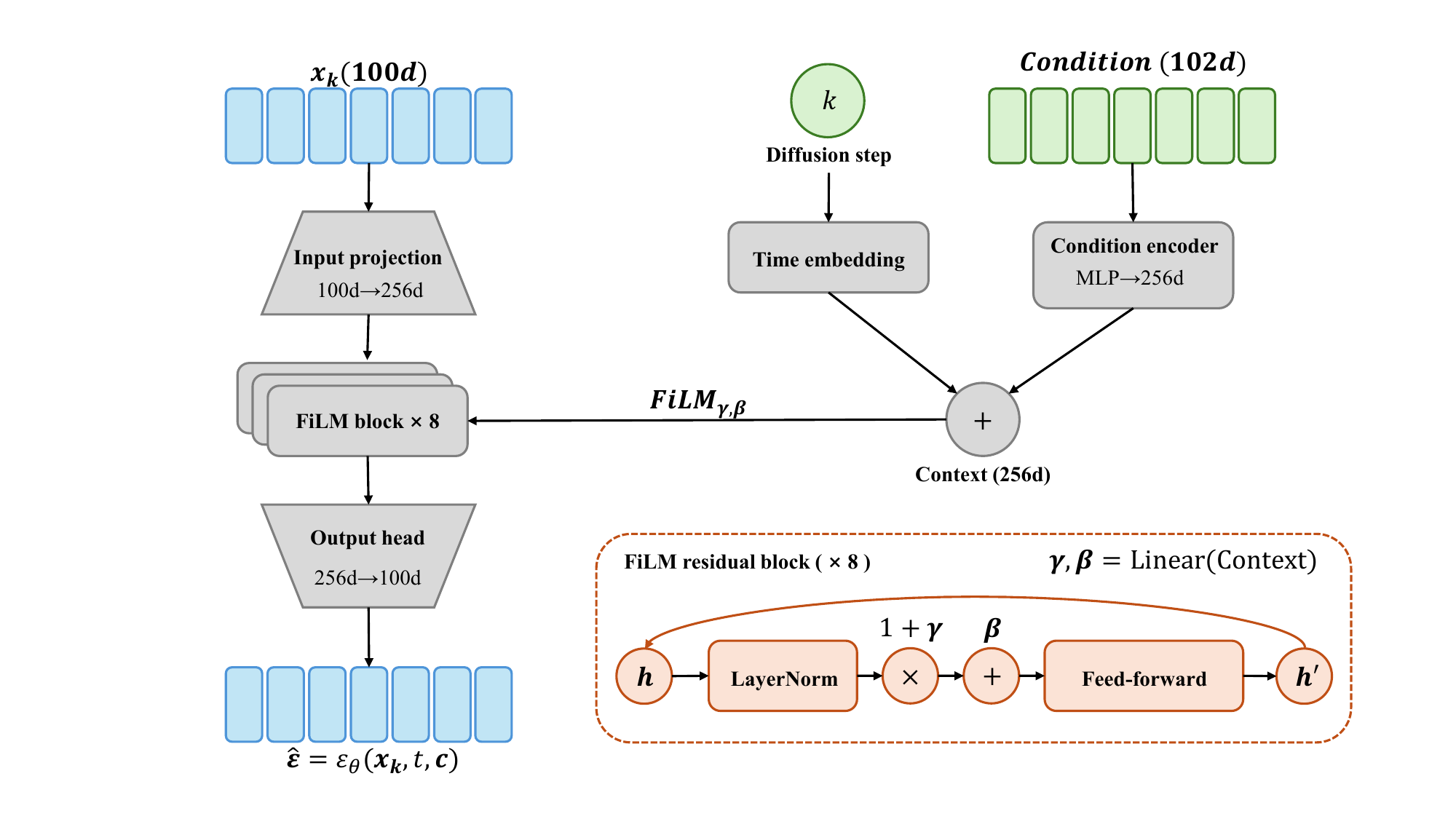}
\caption{Conditional FiLM denoiser architecture.}
\label{fig:model_architecture}
\end{figure}

Figure~\ref{fig:model_architecture} summarizes the architecture. The diffusion step $k$ and the conditioning state $c_t$ are first encoded and concatenated into a context vector. This context is passed through a stack of eight FiLM residual blocks. In each block the context vector produces a pair of modulation parameters, the scale $\gamma_\ell$ and shift $\beta_\ell$, which transform the layer-normalised hidden features before the feed-forward layer. The FiLM mechanism allows the denoiser to adapt its representations to both the current diffusion timestep and the prevailing market conditions. The output of the final residual block is projected through a linear head that predicts the noise $\epsilon$ added to the target in the forward process.

\subsection{Training Objectives}

Training starts from the standard DDPM forward process \cite{Ho2020DDPM,NicholDhariwal2021}
\begin{equation}\label{eq:qsample}
    x_k = \sqrt{\bar{\alpha}_k}\,x_0 + \sqrt{1 - \bar{\alpha}_k}\,\epsilon, \qquad \epsilon \sim \mathcal{N}(0, I),
\end{equation}
where $x_0$ denotes the clean target, $\bar{\alpha}_k$ denotes the cumulative noise-retention coefficient at step $k$, $\epsilon$ denotes Gaussian noise, and $I$ denotes the identity covariance matrix. The noise-matching loss is
\begin{equation}\label{eq:ldiff}
    \mathcal{L}_{\mathrm{mse}} = \mathbb{E}_{x_0, c_t, k, \epsilon} \big\lVert \epsilon - \epsilon_\theta(x_k, k, c_t) \big\rVert_2^2,
\end{equation}
where $\mathcal{L}_{\mathrm{mse}}$ denotes the baseline denoising objective and $\epsilon_\theta(\cdot)$ denotes the neural noise predictor with parameters $\theta$.
To impose financially meaningful structure on generated surfaces, the noised target is first inverted to recover the clean-target estimate,
\begin{equation}\label{eq:x0hat}
    \widehat{x}_0 = \frac{1}{\sqrt{\bar{\alpha}_k}}\,x_k - \sqrt{\frac{1}{\bar{\alpha}_k} - 1}\,\widehat{\epsilon},
\end{equation}
where $\widehat{x}_0$ denotes the reconstructed clean target and $\widehat{\epsilon}$ denotes the estimated noise output by the denoiser, 
whose surface block gives the predicted increment $\widehat{\Delta g}_t$. Adding the previous-day surface reconstructs the predicted level,
\begin{equation}\label{eq:greconstruct}
    \widehat{g}_t = g_{t-1} + \widehat{\Delta g}_t,
\end{equation}
where $\widehat{g}_t$ denotes the reconstructed next-day log-volatility surface and $\widehat{\Delta g}_t$ denotes its predicted increment component. The paper compares four training objectives as follows, and names them $\mathrm{DDPM}_{\mathrm{mse}}$, $\mathrm{DDPM}_{\mathrm{smooth}}$, $\mathrm{DDPM}_{\mathrm{arbitrage}}$, $\mathrm{DDPM}_{\mathrm{hybrid}}$ respectively. 

\paragraph{MSE loss.} We trained a baseline model only use MSE loss, according to Equation~\eqref{eq:ldiff}.

\paragraph{Smoothness penalty.} Following VolGAN
\cite{VuleticCont2024}, the smoothness-regularised variant penalizes
first-order finite differences of the reconstructed log-volatility
surface in the moneyness and maturity directions. The resulting
penalties are discrete Sobolev-type seminorms, consistent with earlier
regularization approaches for volatility-surface reconstruction
\cite{JacksonSuliHowison1999,BenHamidaCont2005}. Let
$\{m_i\}_{i=1}^{N_m}\times\{\tau_j\}_{j=1}^{N_\tau}$ denote the
moneyness--maturity grid. We define
\begin{equation}\label{eq:lm}
    \mathcal{L}_m(\widehat{g}_t)
    =
    \sum_{i=1}^{N_m-1}\sum_{j=1}^{N_\tau}
    \frac{
    \left(
    \widehat{g}_t(m_{i+1},\tau_j)
    -
    \widehat{g}_t(m_i,\tau_j)
    \right)^2
    }{
    \left|m_{i+1}-m_i\right|^2
    }
    \simeq
    \left\|\partial_m\widehat{g}_t\right\|_{L^2}^2,
\end{equation}
and
\begin{equation}\label{eq:ltau}
    \mathcal{L}_\tau(\widehat{g}_t)
    =
    \sum_{i=1}^{N_m}\sum_{j=1}^{N_\tau-1}
    \frac{
    \left(
    \widehat{g}_t(m_i,\tau_{j+1})
    -
    \widehat{g}_t(m_i,\tau_j)
    \right)^2
    }{
    \left|\tau_{j+1}-\tau_j\right|^2
    }
    \simeq
    \left\|\partial_\tau\widehat{g}_t\right\|_{L^2}^2.
\end{equation}
The corresponding contribution to the training objective is
\begin{equation}\label{eq:smooth}
    \mathcal{L}_{\mathrm{smooth}}
    =\mathcal{L}_{\mathrm{mse}}+\lambda_m\mathbb{E}\left[\mathcal{L}_m\right(\widehat{g}_t)]+\lambda_\tau\mathbb{E}\left[\mathcal{L}_\tau\right(\widehat{g}_t)],
\end{equation}
where $\mathbb{E}[\cdot]$ denotes the empirical expectation, namely the arithmetic average of the corresponding per-surface penalty over all reconstructed surfaces in the training mini-batch. 

\paragraph{Arbitrage penalty.} The discrete arbitrage conditions are evaluated on the reconstructed
volatility surface. We therefore define $\widehat{\sigma}_t(\bm{m},\bm{\tau})
    =
    \exp\!\left(
        \widehat{g}_t(\bm{m},\bm{\tau})
    \right).$ The total arbitrage penalty
$\Phi\big(\widehat{\sigma}_t(\bm{m},\bm{\tau})\big)$ is defined in
Equation~\eqref{eq:arbpenalty}, with its calendar, vertical-spread, and
butterfly components given respectively by
Equations~\eqref{eq:p1}--\eqref{eq:p3}. To permit component-specific
regularization strengths, the arbitrage-aware objective is written as
\begin{equation}\label{eq:arbitrage}
\begin{split}
    \mathcal{L}_{\mathrm{arbitrage}}
    &=
    \mathcal{L}_{\mathrm{mse}}
    +
        \lambda_{\mathrm{calendar}}\,
        \mathbb{E}\Big[p_{\mathrm{calendar}}
        \big(
            \widehat{\sigma}_t(\bm{m},\bm{\tau})
        \big)\Big]
        \\
        &\qquad
        +
        \lambda_{\mathrm{spread}}\,
        \mathbb{E}\Big[p_{\mathrm{spread}}
        \big(
            \widehat{\sigma}_t(\bm{m},\bm{\tau})
        \big)\Big]
        \\
        &\qquad
        +
        \lambda_{\mathrm{butterfly}}\,
        \mathbb{E}\Big[p_{\mathrm{butterfly}}
        \big(
            \widehat{\sigma}_t(\bm{m},\bm{\tau})
        \big)\Big].
\end{split}
\end{equation}

\paragraph{Hybrid objective.} The hybrid specification combines the smoothness penalties in Equations~\eqref{eq:smooth} with the arbitrage penalties defined in Equations~\eqref{eq:arbitrage}:
\begin{equation}\label{eq:lc}
\begin{split}
    \mathcal{L}_{\mathrm{hybrid}} &= \mathcal{L}_{\mathrm{mse}} + \lambda_m\mathbb{E}\left[\mathcal{L}_m\right(\widehat{g}_t)]+\lambda_\tau\mathbb{E}\left[\mathcal{L}_\tau\right(\widehat{g}_t)]
    \\
    &\qquad
    +
       \lambda_{\mathrm{calendar}}\,
        \mathbb{E}\Big[p_{\mathrm{calendar}}
        \big(
            \widehat{\sigma}_t(\bm{m},\bm{\tau})
        \big)\Big]
        \\
        &\qquad
        +
        \lambda_{\mathrm{spread}}\,
        \mathbb{E}\Big[p_{\mathrm{spread}}
        \big(
            \widehat{\sigma}_t(\bm{m},\bm{\tau})
        \big)\Big]
        \\
        &\qquad
        +
        \lambda_{\mathrm{butterfly}}\,
        \mathbb{E}\Big[p_{\mathrm{butterfly}}
        \big(
            \widehat{\sigma}_t(\bm{m},\bm{\tau})
        \big)\Big].
\end{split}
\end{equation}

\subsection{Simulation}

\begin{figure}[H]
\centering
\includegraphics[width=0.49\textwidth]{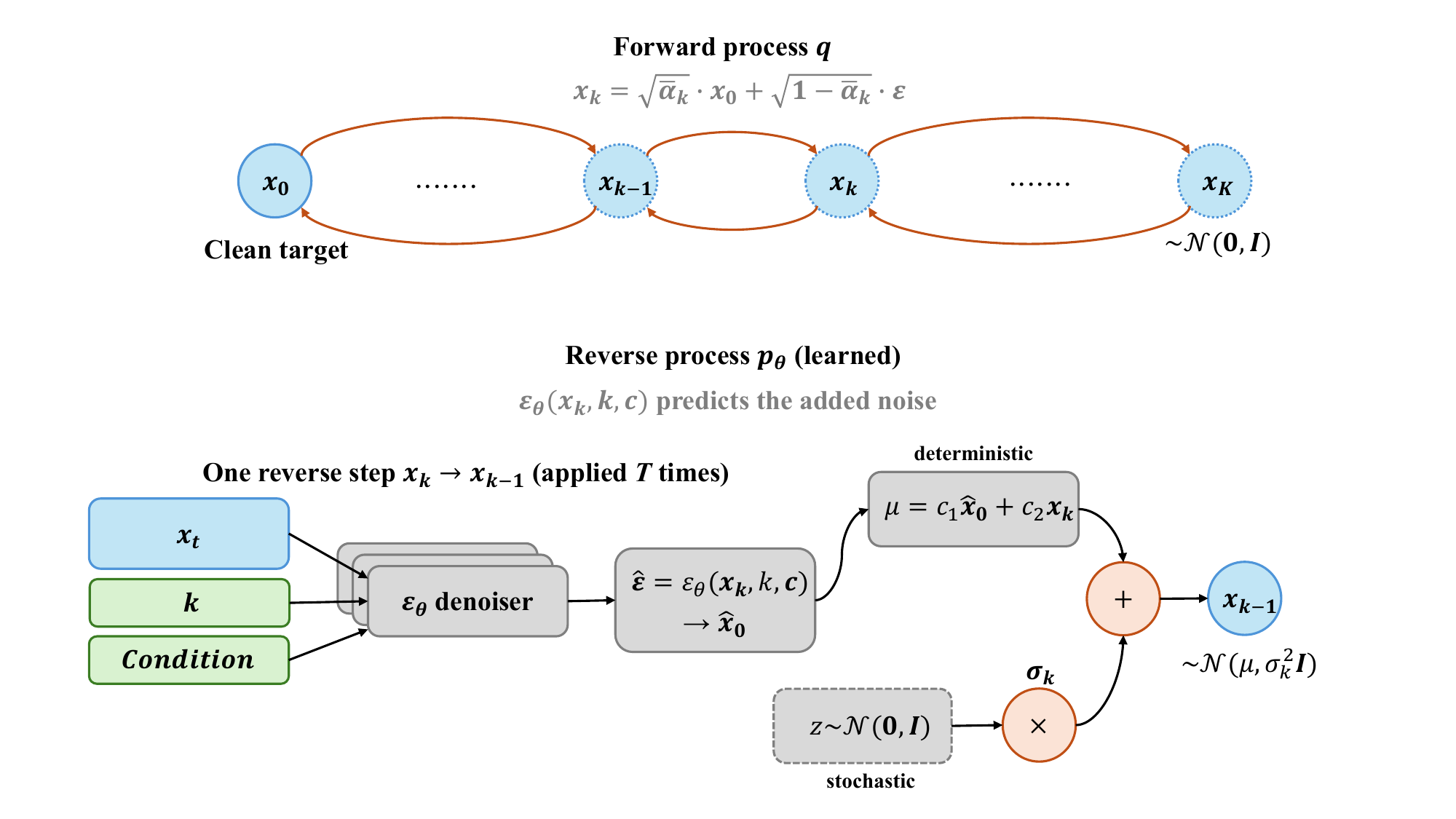}
\caption{Forward noising and learned reverse denoising in the conditional diffusion model.}
\label{fig:diffusion_process}
\end{figure}

Figure~\ref{fig:diffusion_process} illustrates the forward and reverse processes. All penalty coefficients are determined by gradient matching~\cite{VuleticCont2024} to balance the main-loss gradients; for the hybrid model the resulting weights are $\lambda_{\text{calendar}}=0.2739$, $\lambda_{\text{spread}}=0.0671$, $\lambda_{\text{butterfly}}=0.0166$, $\lambda_m=0.0086$, and $\lambda_\tau=0.0011$, and other variants yield similar values. To further reduce arbitrage, we apply scenario re-weighting~\cite{ContVuletic2023,VuleticCont2024} with $\beta=50$, which concentrates mass on low-penalty surfaces while retaining effective sample sizes above 995 out of 1000.

The experimental design, summarized in Table~\ref{tab:setup}.We uses 100 US equities split 50/50 for training and transfer tests, with a chronological 2010--2022/2023--2024 split. All models are trained on pooled in-sample data with AdamW (learning rate $6\times10^{-4}$, batch size 4096, 1000 epochs, EMA 0.999); the VolGAN benchmark is retrained on the same pool with recalibrated smoothness penalties. Z-score normalization parameters are estimated solely on each dimention from the in-sample training set. Each conditioning date yields $S=1000$ scenarios. A single random seed is used for reproducibility.

\begin{table}[H]
\caption{Core empirical design and diffusion configuration.}
\label{tab:setup}
\centering
\small
\begin{tabularx}{\columnwidth}{l>{\raggedright\arraybackslash}X}
\toprule
Item & Setting \\
\midrule
Stocks & 100 US equities \\
Cross-sectional split & Random 50 in-sample, 50 out-of-sample \\
Time split & 2010--2022 train, 2023--2024 test \\
Surface grid & 11 moneyness nodes in $\mathcal{M}$ and 9 maturity nodes in $\mathcal{T}$ \\
Conditioning state & two lagged returns, realized volatility, previous-day log-IVS \\
Target & next-day return and log-IVS increment \\
Denoiser & FiLM MLP, hidden width $256$, 8 residual blocks \\
Diffusion steps & $K=1000$ with cosine schedule \\
Samples per state & $S=1000$ \\
Benchmarks & VolGAN trained with 50 stocks \\
\bottomrule
\end{tabularx}
\end{table}

\section{Results}\label{sec:results}

Evaluation is based on three metrics: static-arbitrage violations of the generated implied volatility surfaces, coverage of the 95\% predictive interval for returns, and PCA-based agreement in relative variance allocation of daily surface increments. Across all three criteria, the diffusion specifications consistently outperform the VolGAN benchmark~\cite{VuleticCont2024}. The MSE-only variant yields the lowest arbitrage penalties, while the arbitrage-aware variant achieves return coverage closest to the nominal target and the best principal-component alignment in most comparisons.

All evaluations use the test-period set, which is split into 50 in-sample stocks (included in training) and 50 out-of-sample stocks (excluded from training). Similar performance across the two groups indicates robust generalization. Throughout this section, tables report in-sample and out-of-sample panels side by side for direct comparison.

\subsection{Arbitrage Violations}

\begin{table*}[t!]
\caption{Arbitrage violation summaries.}
\label{tab:arb_severity}
\centering
\small
\setlength{\tabcolsep}{2.8pt}
\begin{tabular}{l*{10}{r}}
\toprule
& \multicolumn{2}{c}{Mean of $\Phi$}
& \multicolumn{3}{c}{Mean of components ($\beta>0$)}
& \multicolumn{5}{c}{Quantiles of $\Phi$ ($\beta>0$)} \\
\cmidrule(lr){2-3}
\cmidrule(lr){4-6}
\cmidrule(lr){7-11}
Model
& $\beta=0$
& $\beta>0$
& Calendar
& Spread
& Butterfly
& $Q_{0.05}$
& $Q_{0.25}$
& Median
& $Q_{0.75}$
& $Q_{0.95}$ \\
& $(\times 10^{-3})$
& $(\times 10^{-3})$
& $(\times 10^{-6})$
& $(\times 10^{-5})$
& $(\times 10^{-4})$
& $(\times 10^{-6})$
& $(\times 10^{-5})$
& $(\times 10^{-4})$
& $(\times 10^{-3})$
& $(\times 10^{-3})$ \\
\midrule

\multicolumn{11}{l}{\textit{Panel A: In-sample stocks (IS, $n=50$)}} \\
\addlinespace[0.2em]

Data
& 1.38
& 1.38
& 64.60
& 25.40
& 10.60
& 0.09
& 0.01
& 0.46
& 0.42
& 1.90 \\

VolGAN
& 2.90
& 1.80
& 26.40
& 3.41
& 17.40
& 46.00
& 17.94
& 9.72
& 2.10
& 7.30 \\

$\mathrm{DDPM}_{\mathrm{mse}}$
& \underline{\textbf{1.10}}
& \underline{\textbf{0.33}}
& 3.19
& \underline{\textbf{0.04}}
& \underline{\textbf{3.27}}
& \underline{\textbf{4.98}}
& \underline{\textbf{2.98}}
& \underline{\textbf{0.90}}
& \underline{\textbf{0.25}}
& \underline{\textbf{1.10}} \\

$\mathrm{DDPM}_{\mathrm{smooth}}$
& 1.50
& 0.52
& 5.48
& 0.24
& 5.08
& 24.00
& 9.84
& 2.06
& 0.58
& 1.50 \\

$\mathrm{DDPM}_{\mathrm{arbitrage}}$
& \underline{\textbf{1.10}}
& 0.49
& 2.36
& 0.11
& 4.89
& 12.80
& 7.60
& 1.64
& 0.53
& 1.50 \\

$\mathrm{DDPM}_{\mathrm{hybrid}}$
& \underline{\textbf{1.10}}
& 0.48
& \underline{\textbf{1.64}}
& 0.11
& 4.81
& 15.60
& 7.65
& 1.81
& 0.43
& 1.40 \\

\midrule

\multicolumn{11}{l}{\textit{Panel B: Out-of-sample stocks (OOS, $n=50$)}} \\
\addlinespace[0.2em]

Data
& 0.56
& 0.56
& 1.46
& 0.68
& 5.48
& 0.10
& 0.02
& 0.72
& 0.53
& 2.40 \\

VolGAN
& 2.20
& 2.00
& 15.10
& \underline{\textbf{0.01}}
& 19.40
& 35.30
& 20.24
& 5.02
& 1.70
& 8.40 \\

$\mathrm{DDPM}_{\mathrm{mse}}$
& \underline{\textbf{0.51}}
& \underline{\textbf{0.42}}
& 1.60
& 0.06
& \underline{\textbf{4.17}}
& \underline{\textbf{4.91}}
& \underline{\textbf{1.78}}
& \underline{\textbf{0.73}}
& \underline{\textbf{0.54}}
& \underline{\textbf{1.70}} \\

$\mathrm{DDPM}_{\mathrm{smooth}}$
& 0.70
& 0.59
& 5.77
& 0.14
& 5.84
& 13.60
& 6.77
& 1.85
& 0.69
& 2.40 \\

$\mathrm{DDPM}_{\mathrm{arbitrage}}$
& 0.61
& 0.59
& \underline{\textbf{0.79}}
& 0.04
& 5.92
& 9.29
& 4.64
& 1.58
& 0.57
& 2.60 \\

$\mathrm{DDPM}_{\mathrm{hybrid}}$
& 0.63
& 0.61
& 0.80
& 0.09
& 6.09
& 8.87
& 5.49
& 1.71
& 0.57
& 2.90 \\

\midrule

\multicolumn{11}{l}{\textit{Panel C: Full universe ($n=100$)}} \\
\addlinespace[0.2em]

Data
& 0.97
& 0.97
& 33.00
& 13.00
& 8.02
& 0.09
& 0.02
& 0.54
& 0.49
& 2.20 \\

VolGAN
& 2.60
& 1.90
& 20.80
& 1.71
& 18.40
& 41.50
& 18.67
& 6.86
& 1.90
& 7.70 \\

$\mathrm{DDPM}_{\mathrm{mse}}$
& \underline{\textbf{0.81}}
& \underline{\textbf{0.37}}
& 2.39
& \underline{\textbf{0.05}}
& \underline{\textbf{3.72}}
& \underline{\textbf{4.60}}
& \underline{\textbf{2.81}}
& \underline{\textbf{0.88}}
& \underline{\textbf{0.39}}
& \underline{\textbf{1.60}} \\

$\mathrm{DDPM}_{\mathrm{smooth}}$
& 1.10
& 0.55
& 5.63
& 0.19
& 5.46
& 16.30
& 7.73
& 2.01
& 0.65
& 2.30 \\

$\mathrm{DDPM}_{\mathrm{arbitrage}}$
& 0.86
& 0.54
& 1.58
& 0.08
& 5.41
& 10.40
& 5.43
& 1.62
& 0.54
& 2.50 \\

$\mathrm{DDPM}_{\mathrm{hybrid}}$
& 0.88
& 0.55
& \underline{\textbf{1.22}}
& 0.10
& 5.45
& 9.95
& 5.82
& 1.77
& 0.53
& 2.80 \\

\bottomrule
\end{tabular}

\vspace{0.3em}

\begin{minipage}{\textwidth}
\footnotesize
Note: The statistics are computed from stock-level test-period mean arbitrage penalties. For each stock-date and model, the total violation and its components are aggregated over $S=1000$ generated scenarios. The $\beta=0$ column uses equal scenario weights, whereas the $\beta>0$ columns use weights determined by the total arbitrage violation $\Phi$. The same scenario weights are applied when computing the reweighted Calendar, Spread, and Butterfly component means. The reported quantiles are computed under $\beta>0$ and describe the cross-sectional distribution of stock-level test-period mean total penalties within each stock group. The Data row is calculated directly from the observed surfaces and is unaffected by scenario reweighting; its empirical mean is therefore repeated in the two total-mean columns for reference. Data is excluded from the model ranking. All displayed values are rounded to two decimal places after applying the scale factors reported beneath the column headings. Bold entries identify the lowest violation among the generative models based on the unrounded estimates. Because the observed Data row differs substantially between the two stock groups, absolute differences between in-sample and out-of-sample model penalties should not be interpreted as a direct measure of transfer performance.
\end{minipage}
\end{table*}

For model $m$ and stock $i$ the mean arbitrage penalty on test date $t$ under reweighting strength $\beta$ is \begin{equation} \label{eq:scenario_mean_penalty} \widetilde{\Phi}_{i,t}^{m,\beta} = \sum_{s=1}^{S} w_{i,t,s}^{m,\beta} \Phi_{i,t,s}^{m}, \qquad S=1000. \end{equation} The weights sum to one across the generated scenarios. When $\beta=0$ each scenario receives weight $1/S$. This gives the result without scenario reweighting. When $\beta>0$ the scenario weights penalise outputs with larger arbitrage violations. The stock-level test-period mean penalty is \begin{equation} \label{eq:ticker_mean_penalty} \overline{\Phi}_{i}^{m,\beta} = \frac{1}{B_i} \sum_{t=1}^{B_i} \widetilde{\Phi}_{i,t}^{m,\beta}. \end{equation} Here $B_i$ is the number of available test dates for stock $i$. For each stock group the table reports the cross-sectional mean and five quantiles of the stock-level violations. The reported groups are the in-sample stocks, the out-of-sample stocks and the full universe.

Table~\ref{tab:arb_severity} first provides direct evidence for the effectiveness of scenario reweighting. Every generative model has a lower mean arbitrage violation under $\beta>0$ than under $\beta=0$ in all three stock groups. The reduction appears for VolGAN and for every diffusion specification. We can reasonably conclude that scenario reweighting reduces the contribution of high-violation outputs and lowers the aggregate arbitrage violation \cite{ContVuletic2023, VuleticCont2024}.

The comparison focuses on the generative models, excluding the Data row from ranking. Before and after reweighting, every diffusion specification yields lower mean violations than VolGAN across all stock groups, with lower reweighted means and lower violations at all five reported quantiles. The butterfly component dominates total violations for all models. The diffusion family substantially reduces it relative to VolGAN, while calendar penalties are also generally lower and the spread component remains small. Within the diffusion family, the MSE-only variant achieves the lowest total violations, recording the lowest reweighted mean in each group and the lowest value at every reweighted quantile, driven primarily by the lowest butterfly violation in all three groups. This advantage persists equally for out-of-sample stocks. Every diffusion variant continues to outperform VolGAN, with the MSE-only model remaining best, confirming that the lower arbitrage violations reflect a genuine transferable capability rather than overfitting to the in-sample stocks.

\subsection{Stock Return Risk Prediction}

\begin{table*}[t]
\caption{Coverage of the 95\% predictive intervals.}
\label{tab:return_calibration}
\centering
\begin{tabular}{lrrr}
\toprule
Model
& In-sample stocks
& Out-of-sample stocks
& Full universe
\\

& Coverage (\%)
& Coverage (\%)
& Coverage (\%)
\\
\midrule

\multicolumn{4}{l}{%
    \textit{Panel A: without scenario reweighting ($\beta=0$)}%
}
\\
\addlinespace[0.2em]

VolGAN
& 73.50
& 76.62
& 75.06
\\

$\mathrm{DDPM}_{\mathrm{mse}}$
& 93.85
& 94.00
& 93.92
\\

$\mathrm{DDPM}_{\mathrm{smooth}}$
& 96.01
& 96.01
& 96.01
\\

$\mathrm{DDPM}_{\mathrm{arbitrage}}$
& \underline{\textbf{94.38}}
& \underline{\textbf{94.53}}
& \underline{\textbf{94.45}}
\\

$\mathrm{DDPM}_{\mathrm{hybrid}}$
& 95.62
& 95.72
& 95.67
\\
\midrule
\multicolumn{4}{l}{%
    \textit{Panel B: with scenario reweighting ($\beta>0$)}%
}
\\
\addlinespace[0.2em]

VolGAN
& 73.32
& 76.33
& 74.83
\\

$\mathrm{DDPM}_{\mathrm{mse}}$
& 94.03
& 94.11
& 94.07
\\

$\mathrm{DDPM}_{\mathrm{smooth}}$
& 96.04
& 96.08
& 96.06
\\

$\mathrm{DDPM}_{\mathrm{arbitrage}}$
& \underline{\textbf{94.50}}
& \underline{\textbf{94.60}}
& \underline{\textbf{94.55}}
\\

$\mathrm{DDPM}_{\mathrm{hybrid}}$
& 95.74
& 95.76
& 95.75
\\

\bottomrule
\end{tabular}

\vspace{0.3em}

\begin{minipage}{\linewidth}
\footnotesize
Note: Coverage is computed separately for each stock and then averaged with equal weight across stocks.
The in-sample and out-of-sample groups each contain 50 stocks, while the full sample contains 100 stocks.
Bold entries denote the values closest to the nominal 95\% target. Panel A computes the predictive quantiles from the unweighted generated scenarios. Panel B computes the predictive quantiles after scenario reweighting. Each generated return is associated with the same scenario weight as its corresponding IVS.
\end{minipage}
\end{table*}

\begin{table*}[t]
\centering
\caption{
Cross-stock mean explained variance ratios of the first three principal components under one-step conditional generation.
}
\label{tab:mean_pca_cross_model}
\begin{tabular}{l *{9}{>{\centering\arraybackslash}p{2em}}}
\toprule
& \multicolumn{3}{c}{In-sample (\%)}
& \multicolumn{3}{c}{Out-of-sample (\%)}
& \multicolumn{3}{c}{Full-universe (\%)} \\

\cmidrule(lr){2-4}
\cmidrule(lr){5-7}
\cmidrule(lr){8-10}
Model
& PC1 & PC2 & PC3
& PC1 & PC2 & PC3
& PC1 & PC2 & PC3 \\
\midrule

Data
& 81.62 & 7.19 & 3.77
& 80.62 & 7.52 & 3.96
& 81.12 & 7.35 & 3.87 \\

VolGAN
& 76.41 & 9.18 & 6.54
& 76.87 & 8.73 & 6.73
& 76.64 & 8.95 & 6.64 \\

$\mathrm{DDPM}_{\mathrm{mse}}$
& 79.46 & \underline{\textbf{7.63}} & 4.72
& \underline{\textbf{79.59}} & \underline{\textbf{7.25}} & 4.76
& \underline{\textbf{79.53}} & \underline{\textbf{7.44}} & 4.74 \\

$\mathrm{DDPM}_{\mathrm{smooth}}$
& 83.31 & 6.38 & \underline{\textbf{3.63}}
& 83.66 & 5.89 & \underline{\textbf{3.66}}
& 83.48 & 6.14 & \underline{\textbf{3.64}} \\

$\mathrm{DDPM}_{\mathrm{arbitrage}}$
& \underline{\textbf{82.45}} & \underline{\textbf{6.74}} & \underline{\textbf{3.98}}
& 83.14 & 6.15 & \underline{\textbf{3.99}}
& 82.80 & \underline{\textbf{6.45}} & \underline{\textbf{3.98}} \\

$\mathrm{DDPM}_{\mathrm{hybrid}}$
& \underline{\textbf{80.38}} & 8.86 & 4.32
& \underline{\textbf{80.57}} & \underline{\textbf{8.58}} & 4.40
& \underline{\textbf{80.47}} & 8.72 & 4.36 \\
\bottomrule
\end{tabular}

\vspace{0.3em}

\begin{minipage}{\textwidth}
\footnotesize
Note: The table reports the cross-stock mean explained variance ratio for each principal component. The Data row reports the mean explained variance ratio calculated from the observed data, while each model row reports the corresponding mean implied by the generated scenarios. All values are expressed as percentages. Closeness is measured by the absolute difference between each model value and the corresponding Data value, and the first two closest model values in each column are shown in bold and underlined. The in-sample and out-of-sample groups each contain 50 stocks, while the full universe contains 100 stocks.
\end{minipage}
\end{table*}

For model $m$ and stock $i$, return risk calibration under reweighting setting $\beta$ is measured by the stock-level empirical coverage of the nominal $95\%$ predictive interval \begin{equation} \label{eq:ticker_return_coverage} C_i^{(m,\beta)} = \frac{1}{B_i} \sum_{t=1}^{B_i} \mathbf{1} \left[ q_{2.5\%,i,t}^{(m,\beta)} \leq r_{i,t}^{\mathrm{real}} \leq q_{97.5\%,i,t}^{(m,\beta)} \right],
\end{equation}
where $B_i$ denotes the number of test dates available for stock $i$. $\mathbf{1}\{\cdot\}$ equals $1$ when the condition inside the brackets is satisfied and $0$ otherwise. The quantities $q_{2.5\%,i,t}^{(m,\beta)}$ and $q_{97.5\%,i,t}^{(m,\beta)}$ are the lower and upper predictive quantiles under reweighting setting $\beta$. The quantity $r_{i,t}^{\mathrm{real}}$ is the realized return.

For stock group $G$ with $n_G$ stocks the equally weighted mean coverage is \begin{equation} \label{eq:group_return_coverage} \overline{C}^{G,m,\beta} = \frac{1}{n_G} \sum_{i \in G} C_i^{(m,\beta)}. \end{equation} 
A model that is well calibrated under this criterion should have mean coverage close to the nominal target of $0.95$.

Table~\ref{tab:return_calibration} reports the equally weighted mean coverage of the 95\% predictive intervals across in‑sample, out‑of‑sample, and full‑universe stocks, with Panel A showing unweighted results and Panel B showing results under scenario reweighting . All four diffusion specifications achieve coverage close to the nominal 95\% in every stock group. The near nominal coverage in Panel A shows that the raw diffusion generators already capture realized return risk effectively. Scenario reweighting is therefore not required to obtain well-calibrated return coverage. The coverage is also similar for in‑sample and out‑of‑sample stocks, indicating robust generalization with no evidence of overfitting.

In contrast, VolGAN exhibits severe undercoverage, with its predictive intervals containing the realized return only about 75\% of the time, systematically understating tail risk. Within the diffusion family, the arbitrage‑aware variant achieves the strongest calibration, its mean coverage being closest to the nominal target in all groups.

The persistent performance gap between the diffusion family and VolGAN demonstrates that the generative mechanism itself is highly consequential for probabilistic risk forecasting. Among the diffusion variants, the arbitrage‑aware loss provides the best return‑risk calibration, and the results confirm that explicit arbitrage regularization helps maintain reliable predictive interval coverage for return risk.

\subsection{Principal Component Analysis}

\begin{figure*}[t]
\centering
\includegraphics[width=\textwidth]
{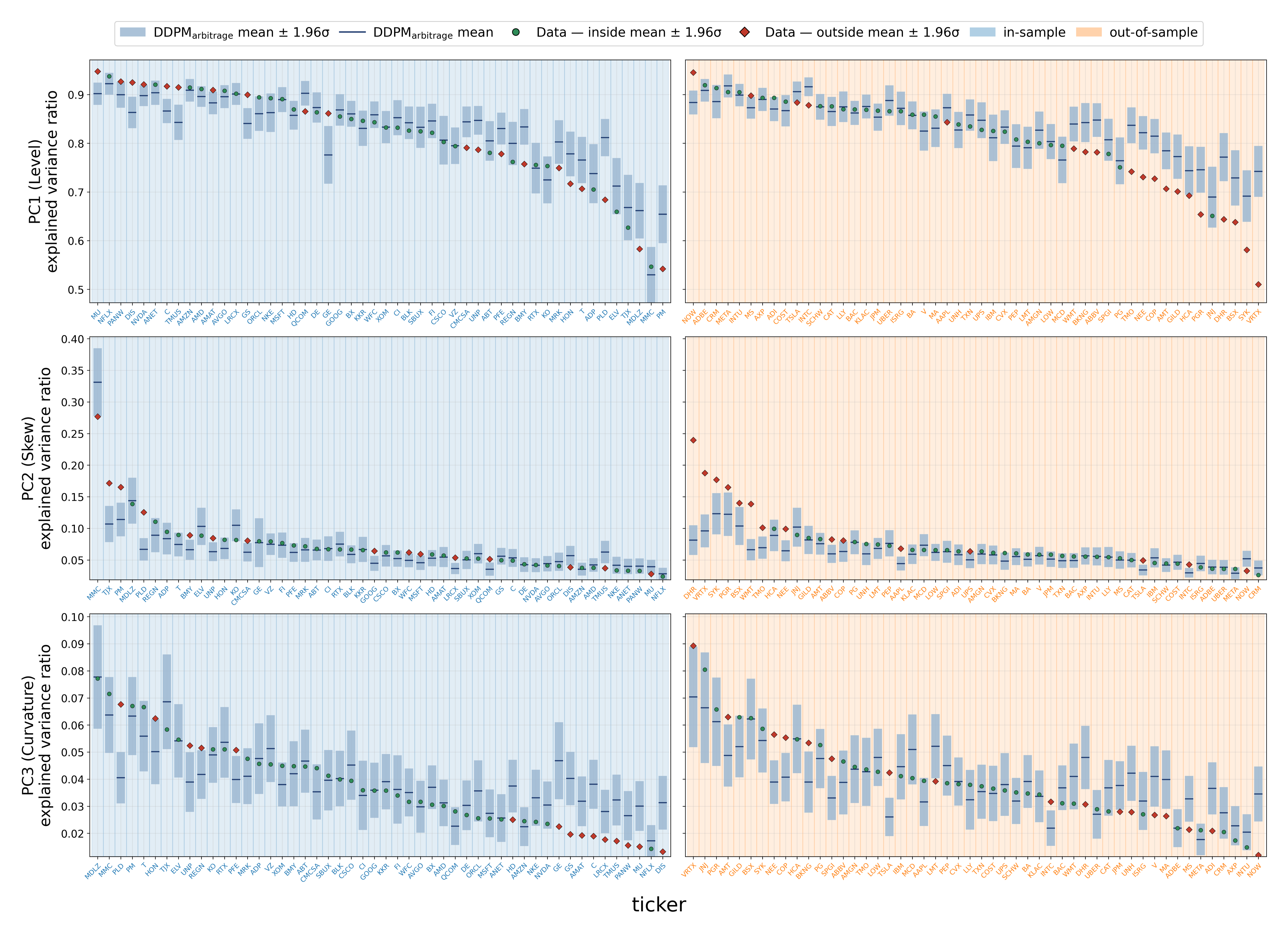}
\caption{
Stock-level comparison of observed and generated explained variance ratios for $\mathrm{DDPM}_{\mathrm{arbitrage}}$ under one-step conditional generation. The top, middle, and bottom rows report PC1 (level), PC2 (skew), and PC3 (curvature), respectively. The left column contains the 50 in-sample stocks, and the right column contains the 50 out-of-sample stocks. For each stock and component, the shaded bar reports the generated average $\pm 1.96$ standard deviations across 1000 one-step conditional scenario sequences constructed along the realized test-period condition sequence. The horizontal line inside each bar marks the generated mean. A green dot indicates that the observed explained variance ratio lies inside the model-implied interval, whereas a red diamond indicates that it lies outside.
}
\label{fig:pca_ticker_panels}
\end{figure*}

For each stock, PCA is applied to the daily log-IVS increments. The explained variance ratio of component $k$ is
\begin{equation}
\label{eq:pca_explained_variance}
    \rho_{i,k} = \frac{\lambda_{i,k}}{\sum_{j=1}^{99}\lambda_{i,j}},
\end{equation}
where $\lambda_{i,k}$ is the $k$th ordered eigenvalue. Figure~\ref{fig:pca_ticker_panels} shows the stock-level comparison for $\mathrm{DDPM}_{\mathrm{arbitrage}}$ under one-step conditional generation. For each stock and component, the generated interval (mean $\pm 1.96$ standard deviations over 1000 scenarios) captures the observed explained variance ratio in the majority of cases, and the model tracks the cross-sectional variation across both in-sample and out-of-sample stocks.

Table~\ref{tab:mean_pca_cross_model} extends the comparison to all models via cross-stock mean explained variance ratios. In the observed data, PC1 accounts for about 81\% of total variation, while PC2 and PC3 account for approximately 7\% and 4\%, respectively. VolGAN underestimates the PC1 share and overestimates PC2 and PC3, while the diffusion models reproduce the allocation more faithfully. $\mathrm{DDPM}_{\mathrm{arbitrage}}$ best matches the PC1 and PC3 means, while $\mathrm{DDPM}_{\mathrm{mse}}$ is closest for PC2. As with the previous criteria, the diffusion family exhibits consistent performance across in-sample and out-of-sample stocks.

\begin{figure}[t]
\centering
\includegraphics[width=0.45\textwidth]
{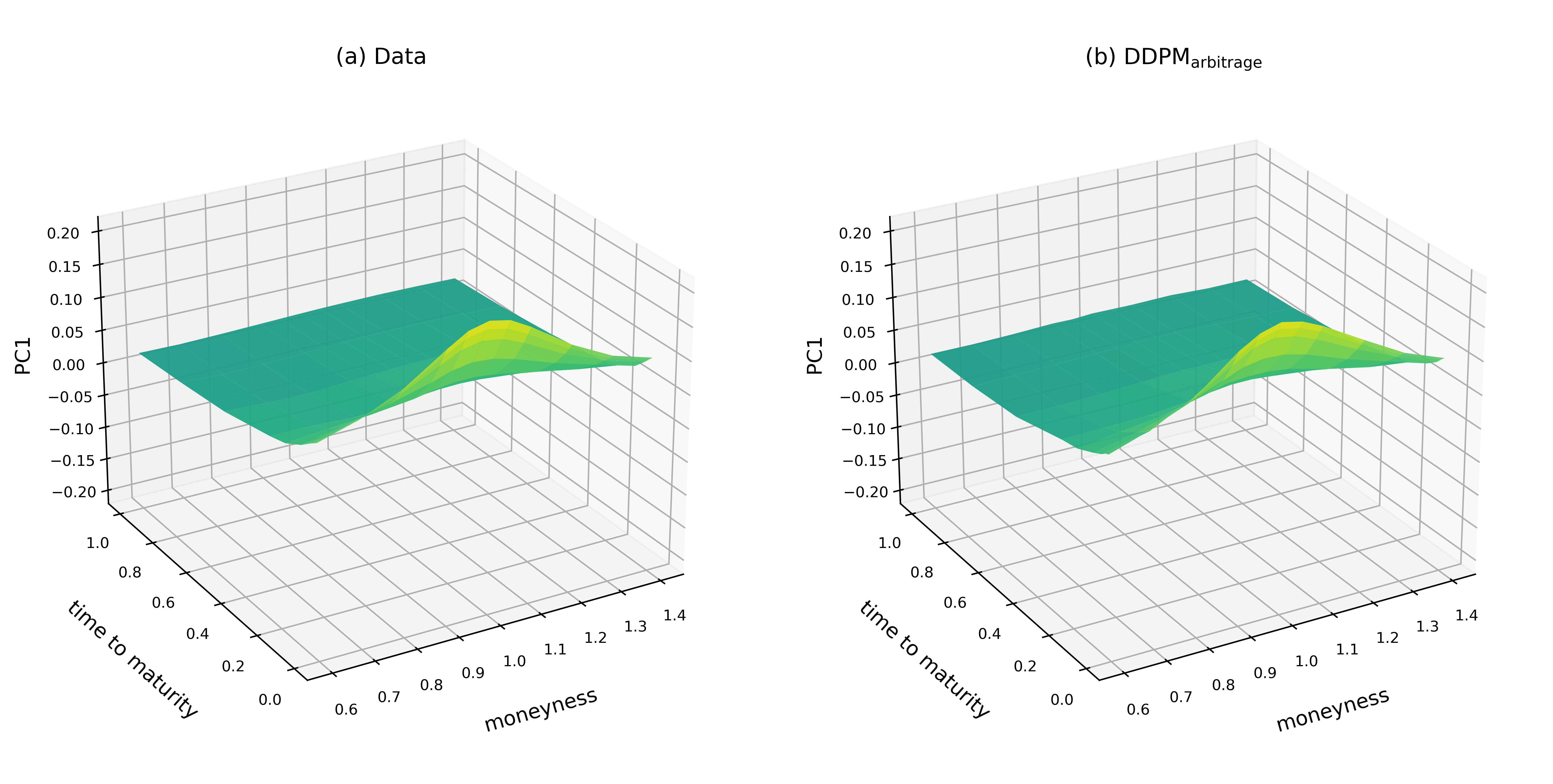}\\[0.4em]
\includegraphics[width=0.45\textwidth]
{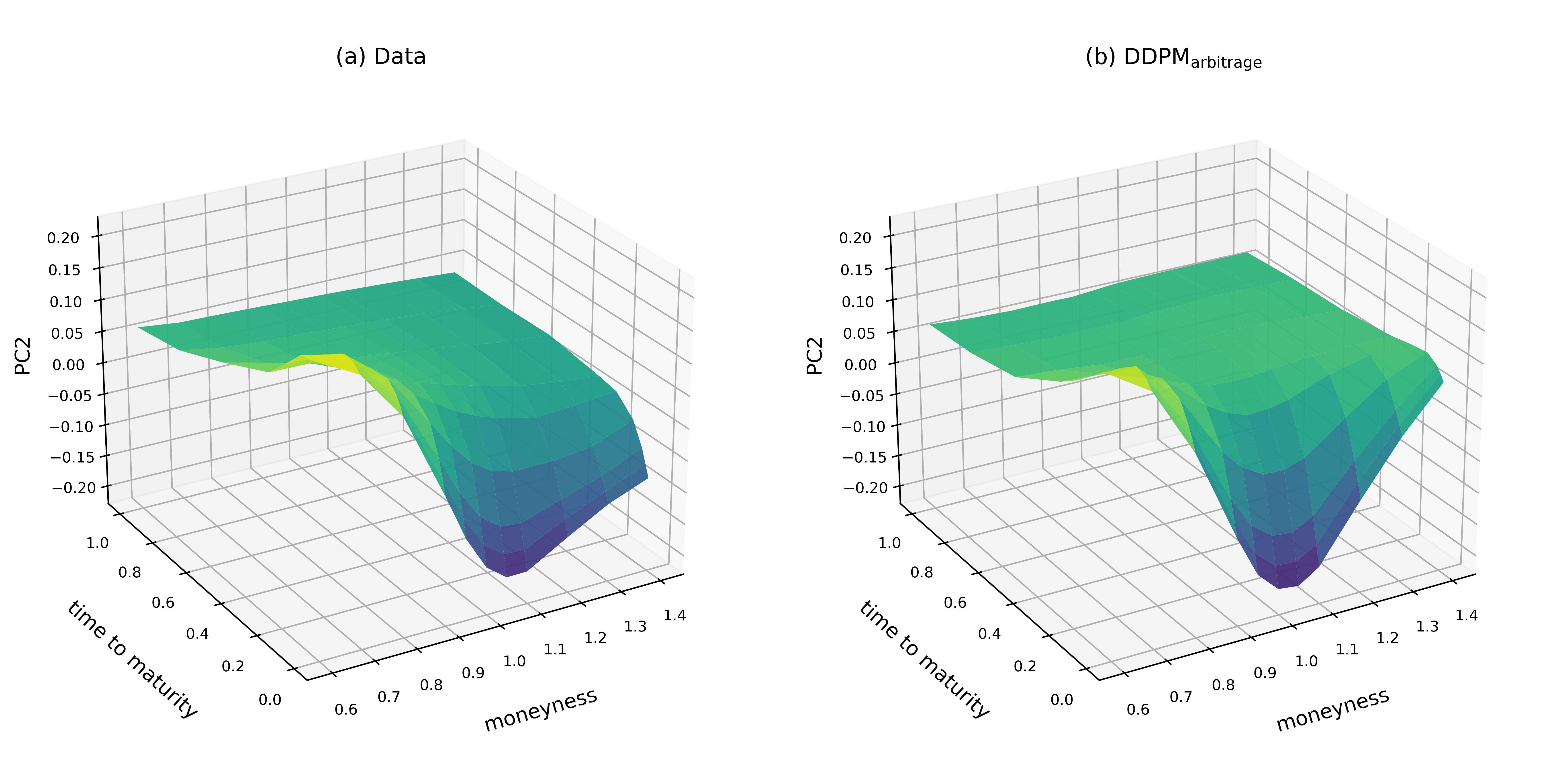}\\[0.4em]
\includegraphics[width=0.45\textwidth]
{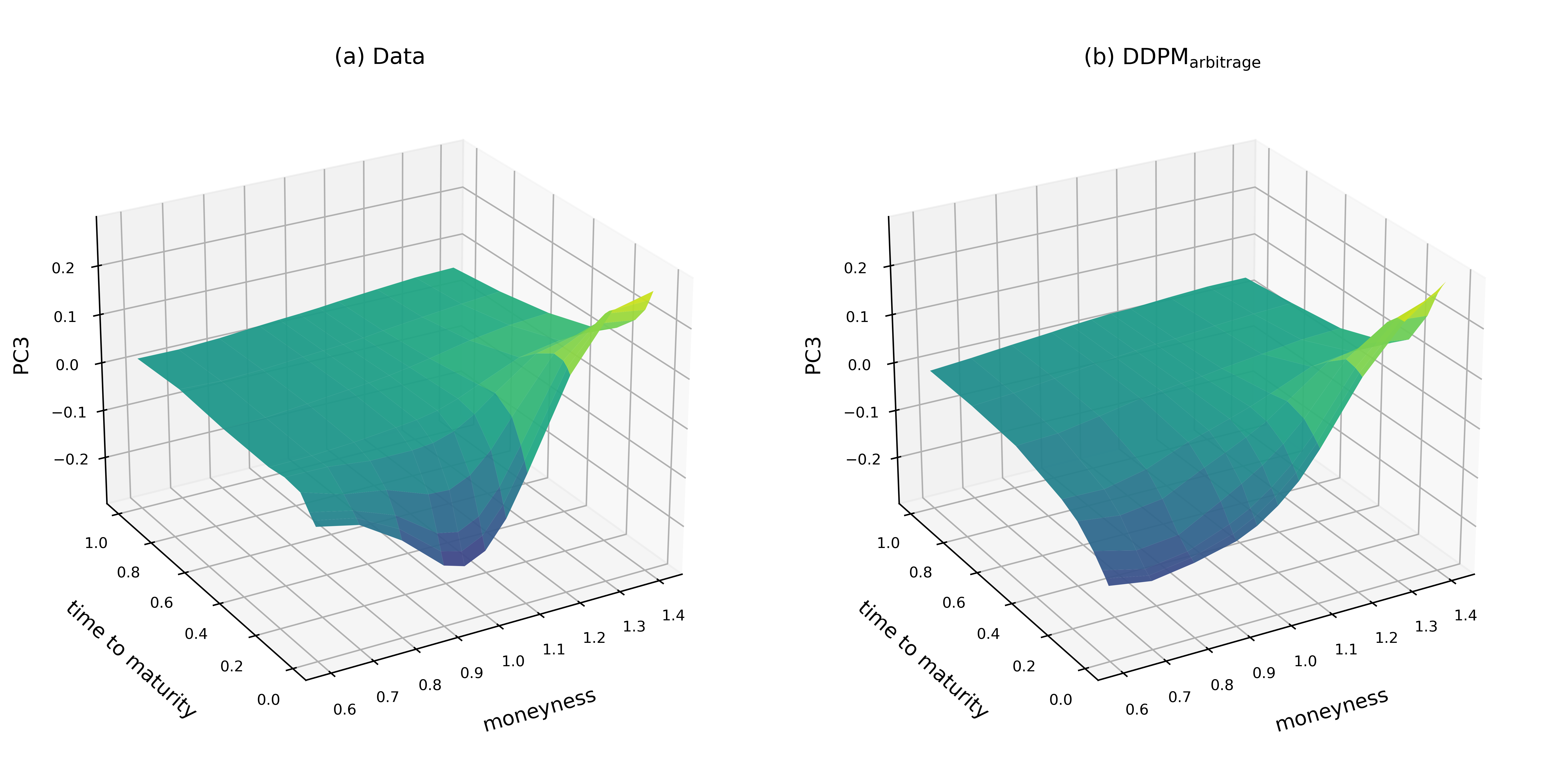}
\caption{
Principal-component loading surfaces for AAPL over the test period. The observed loading surfaces are shown in the left column, and the loading surfaces generated by $\mathrm{DDPM}_{\mathrm{arbitrage}}$ are shown in the right column. The rows report PC1, PC2, and PC3. For each of the 1000 one-step conditional scenario sequences, the generated eigenvectors are sign-aligned with the corresponding observed eigenvectors.
}
\label{fig:pca_aapl_panels}
\end{figure}

Figure~\ref{fig:pca_aapl_panels} complements this analysis by comparing the spatial shapes of the loading surfaces for AAPL. The generated PC1, PC2, and PC3 loading surfaces closely reproduce the observed level, skew, and curvature patterns, confirming that the model captures not only the variance allocation but also the structural co-movement modes of one-day surface increments.

\section{Discussion}\label{sec:discussion}

The core contribution is a dynamic formulation that learns the conditional joint distribution of next-day return and IVS increment, evaluated through static arbitrage penalties, return coverage, and PCA explained variance ratios. Across all three criteria the diffusion family consistently outperforms VolGAN. Notably, no single regularizer dominates: smoothness controls local variation but not arbitrage constraints; explicit arbitrage terms yield the best return calibration and PCA fit, yet the MSE‑only variant achieves the lowest test‑period arbitrage violations. This counterintuitive result likely reflects the smoothed data already lying near an arbitrage‑free manifold, possible gradient interference between the nonlinear penalty and the denoising objective, and the mismatch between training‑time clean‑target penalties and post‑sampling violations. The trade‑off suggests that the penalty–denoiser interaction is complex and warrants component‑wise, multi‑seed investigation.

The pooled multi‑stock results show that all diffusion models outperform VolGAN on both in‑sample and out‑of‑sample stocks, with return coverage and PCA ratios remaining stable across groups, indicating robust transfer without retraining. While out‑of‑sample stocks exhibit a lower empirical arbitrage baseline, the consistent ranking confirms that performance does not collapse under zero‑shot application and that the diffusion advantage is not driven by overfitting.

Important limitations remain: the framework addresses only one‑day‑ahead generation on a fixed grid, enforces static rather than dynamic no‑arbitrage conditions, and assesses calibration solely via a single nominal coverage level. Hence the primary conclusion is that universal conditional diffusion provides a credible, empirically stronger framework for overnight IVS scenario generation. Future work should extend generation horizons, incorporate dynamic arbitrage constraints, evaluate full distributional calibration, normalize penalties by stock‑specific baselines, and systematically analyze how financial penalties interact with diffusion training across multiple random seeds.

\section{Conclusion}\label{sec:conclusion}

We introduced a universal conditional diffusion framework that jointly models next-day IVS dynamics and underlying stock returns, and evaluated it on static-arbitrage violations, return risk coverage, and PCA explained variance ratios. All four diffusion variants outperform the VolGAN benchmark~\cite{VuleticCont2024} on these metrics: the MSE-only specification yields the lowest arbitrage penalties, while the arbitrage-aware variant provides return coverage closest to the nominal 95\% level, and the diffusion family generally produces PCA mean ratios nearer to observed values. The finding that explicit arbitrage regularization does not produce the lowest test-period violations highlights a non-trivial interaction between financial penalties and the denoising objective, motivating further study. Crucially, the diffusion advantage holds for both in-sample and out-of-sample stocks, confirming that the model learns shared dynamics that transfer without retraining. Scenario reweighting further reduces arbitrage violations with minimal impact on return coverage, and together the evidence establishes universal conditional diffusion as a credible, empirically stronger alternative to adversarial IVS generation.

\bibliographystyle{ACM-Reference-Format}
\bibliography{references}

\appendix

\end{document}